\documentclass[aps,prl,twocolumn,showpacs,superscriptaddress,nofootinbib]{revtex4-1}  
\usepackage{graphicx,multirow,subfig}
\usepackage{float}
\usepackage{bm}
\usepackage{braket}
\usepackage{amsmath}
\usepackage{amssymb}
\usepackage{amscd}
\usepackage{latexsym}
\usepackage{slashed}
\usepackage{color}
\usepackage{graphicx}
\usepackage[normalem]{ulem}
\usepackage{color}
\definecolor{orange}{cmyk}{0,0.5,1,0}
\usepackage{verbatim}
\usepackage{rotating}
\usepackage{diagbox}
\usepackage{cancel}
\usepackage[utf8]{inputenc}
\usepackage{array}
\usepackage{caption}

\newcommand{\lagr}{\mathcal{L}}

\def\lsim{\raise0.3ex\hbox{$\;<$\kern-0.75em\raise-1.1ex\hbox{$\sim\;$}}}
\def\gsim{\raise0.3ex\hbox{$\;>$\kern-0.75em\raise-1.1ex\hbox{$\sim\;$}}}

\newcolumntype{L}[1]{>{\raggedright\let\newline\\\arraybackslash\hspace{0pt}}m{#1}}
\newcolumntype{C}[1]{>{\centering\let\newline\\\arraybackslash\hspace{0pt}}m{#1}}
\newcolumntype{R}[1]{>{\raggedleft\let\newline\\\arraybackslash\hspace{0pt}}m{#1}}

\def\be{\begin{equation}}
\def\ee{\end{equation}}
\def\bea{\begin{eqnarray}}
\def\eea{\end{eqnarray}}
\def\nn{\nonumber}

\newcommand{\xdownarrow}[1]{%
	{\left\downarrow\vbox to #1{}\right.\kern-\nulldelimiterspace}
}

\begin{document}
\title{Atomki Anomaly in Family-dependent $U(1)'$ Extension of the Standard Model}

\author{Luigi Delle Rose}
\affiliation{\small INFN, Sezione di Firenze, and Dipartimento di Fisica ed Astronomia,	Universit\`a di  Firenze, Via G. Sansone 1, 50019 Sesto Fiorentino, Italy}
\affiliation{\small School of Physics and Astronomy, University of Southampton,
	Southampton, SO17 1BJ, United Kingdom}

\author{Shaaban Khalil}
\affiliation{Center for Fundamental Physics, Zewail City of Science and Technology, 6 October City, Giza 12588, Egypt}

\author{Simon J.D. King}
\affiliation{\small School of Physics and Astronomy, University of Southampton,
	Southampton, SO17 1BJ, United Kingdom}
\affiliation{INFN, Sezione di Padova, and Dipartimento di Fisica ed Astronomia G. Galilei, Universit\`a di Padova, 
Via Marzolo 8, 35131 Padova, Italy}

\author{Stefano Moretti}
\affiliation{\small School of Physics and Astronomy, University of Southampton,
	Southampton, SO17 1BJ, United Kingdom}

\author{Ahmed M. Thabt}
\affiliation{Center for Fundamental Physics, Zewail City of Science and Technology, 6 October City, Giza 12588, Egypt}

\date{\today}
\begin{abstract}
{In the context of a gauge invariant, non-anomalous and 
family-dependent (non-universal) $U(1)'$ extension of the Standard Model, wherein a new high scale mechanism generates masses and couplings for the first two fermion generations and the standard Higgs mechanism does so for the third one, 
we find solutions to the anomaly observed by the Atomki collaboration in the decay of excited states of Beryllium, in the form of a very light $Z'$ state, stemming from the $U(1)'$ symmetry breaking, with significant axial couplings so as to evade a variety of low scale experimental constraints.}
	\end{abstract}
\maketitle

The Atomki collaboration \cite{Gulyas:2015mia} has recently detected hints of a new light bosonic state, with mass $\simeq 17$ MeV,  from the measurement of the angle between $e^+e^-$ pairs and their invariant mass produced by the 18.15 MeV nuclear transition in the excited state $^8$Be$^*$ \cite{Krasznahorkay:2015iga}\footnote{In fact, also the 17.64 MeV transition eventually appeared to present a similar anomaly, albeit less significant, with a boson mass broadly compatible with the previous one, however, it should be mentioned that this was never documented in a published paper, only in proceeding contributions, so we do not consider it here.} (see also \cite{Krasznahorkay:2017gwn,Krasznahorkay:2017bwh,Krasznahorkay:2017qfd,Krasznahorkay:2018snd}). There have been several studies \cite{Feng:2016jff,Feng:2016ysn,Gu:2016ege,Chen:2016dhm,Liang:2016ffe,Jia:2016uxs,Kitahara:2016zyb,Chen:2016tdz,Seto:2016pks,Neves:2016ugb,Chiang:2016cyf,Ellwanger:2016wfe,DelleRose:2017xil}
trying to explain the nature of this new state which mostly focus on a vector boson solution. In this work we further consider this possible scenario in the context of a rather minimal model: specifically, by extending the Standard Model (SM) with a single family-dependent (non-universal) $U(1)'$ group. 

As the model contains two Abelian groups, $U(1)_Y \times U(1)'$,  there will be a mixing between the hypercharge gauge boson $\hat{B}_\mu$ of the SM and the new $U(1)'$ gauge boson $\hat{B}'_\mu$.  Therefore, the kinetic Lagrangian is given by
\begin{equation}
\mathcal{L}_\mathrm{kin} = - \frac{1}{4} \hat F_{\mu\nu} \hat F^{\mu\nu} - \frac{1}{4} \hat F'_{\mu\nu} \hat F^{'\mu\nu} - \frac{\kappa}{2} \hat F'_{\mu\nu} \hat F^{\mu\nu},
\end{equation}
where $\kappa$ parameterises the level of mixing between the two fields. One may diagonalise the kinetic Lagrangian by a rotation and rescaling of these fields, which leaves the covariant derivative as
\begin{equation}
{\cal D}_\mu = \partial_\mu + .... + i g_1 Y B_\mu + i (\tilde{g} Y + g' z) B'_\mu, 
\end{equation}
where $Y$ and $g_1$ are the hypercharge and its gauge coupling, $z$ and $g'$ are the $U(1)'$ charge and its gauge coupling and $\tilde{g}$ is the mixed gauge coupling between the groups. We break the $U(1)'$ with a new SM-singlet scalar, $\chi$, with a charge $z_\chi$ under the new gauge group, with a Vacuum Expectation Value (VEV) $\braket{\chi}=v'$ inducing a mass term $m_{B'}=g' z_\chi v'$. {It is interesting to note that, for $g' \sim {\cal O}(10^{-4} - 10^{-5})$, as required by several experimental constraints, then $m_{B'}$ can be of order ${\cal O}(10)$ MeV if $v'$, the scale of $U(1)'$ symmetry breaking, is of order ${\cal O}(100-1000)$ GeV.}

This massive vector boson interacts with the SM fermions through the gauge current
\begin{equation}
J^\mu_{Z'} = \sum_f \bar \psi_f \gamma^\mu \left( C_{f, L} P_L + C_{f, R} P_R \right) \psi_f ,
\end{equation}
with Left (L) and Right (R) handed coefficients \cite{DelleRose:2017xil}
\begin{eqnarray}
C_{f,L} &=&  - g_Z s' \left( T^3_f - s_W^2 Q_f \right) + ( \tilde g Y_{f, L} + g' z_{f, L})  c', \nn \\
C_{f,R} &=&  g_Z s_W^2 s' Q_f + ( \tilde g Y_{f, R} + g' z_{f, R}) \, c',
\end{eqnarray}
where we have defined $g_Z = \sqrt{g_1^2 + g_2^2}$ (the Electro-Weak (EW) coupling), $s_W \equiv \sin(\theta _W)$, $c_W \equiv \cos (\theta _W )$, $s' \equiv \sin(\theta')$ and $c' \equiv \cos (\theta ' )$, with $\theta'$ being the angle parameterising the aforementioned gauge mixing and we have also introduced $T_f ^3$ and $Q_f$, the weak isospin and electric charge of the fermion $f$, respectively. Finally, $Y_{f,L/R}$ and $z_{f,L/R}$ represent the hypercharge and $U(1)'$ quantum numbers of the L/R handed fermion. 
By diagonalising the mass matrix of neutral gauge bosons, one finds this mixing angle, $\theta '$, effectively between the SM $Z$ and the new $Z'$ (associated to $U(1)'$), as \cite{Coriano:2015sea}
\bea
\label{eq:mixing2Higgs}
\tan 2 \theta' =  \frac{2  \, g_H \, g_Z}{ g_{H^2} + 4 m_{B'}^2/v^2 - g_Z^2},
\eea
where $g_H= \tilde{g} + 2g'z_H$. \\
We now define the usual Vector (V) and Axial (A) coefficients in the limit of  small gauge coupling and mixing, $g',\tilde{g} \ll 1$,
\begin{eqnarray}
C_{f, V} &=& \frac{C_{f,R} + C_{f,L}}{2}\nonumber\\
	&\simeq &\tilde g  c_W^2 \, Q_f + g' \left[ z_H (T^3_f - 2 s_W^2 Q_f)  + z_{f,V} \right], \\
C_{f, A} &=& \frac{C_{f,R} - C_{f,L}}{2} \simeq  g' \left[   -  z_H \, T^3_f  +   z_{f,A} \right],
\label{eq:Axial}
\end{eqnarray}
where we use the convention $Y_f = Q_f - T_f ^3$, and define the V and A quantum numbers under the $U(1)'$ group, $z_{f,V/A} = 1/2 (z_{f,R} \pm z_{f,L})$.

The Yukawa sector of the SM for quarks and leptons takes the form 
\be 
- \lagr _\mathrm{Yuk} = Y_u \bar{Q} \tilde{H} u_R + Y_d \bar{Q} H d_R + Y_e \bar{L} H e_R.
\label{Yukawa}
\ee
Because of gauge invariance this imposes a condition on the combination of charges of the fields under the $U(1)'$ group:
\begin{equation}
-z_Q -z_H +z_u = -z_Q + z_H +z_d = -z_L + z_H + z_e = 0.
\label{eq:gauge_invariance}
\end{equation}
Inserting these relations into  Eq. (\ref{eq:Axial}), one finds no A couplings to the $Z'$ for quarks and leptons, i.e. $C_{(q,l^\pm),A}\simeq 0$, at leading order in the gauge coupling $g'$.

It is difficult to construct a model (with minimal extra particle content) with only V interactions of fermions to the $Z'$, as opposed to A, because relatively larger couplings\footnote{Though still in the regime $(g,\tilde{g}) \ll 1$.} are required to achieve a successfully high rate for the transition ${}^8 \textrm{Be}^* \rightarrow {}^8 \textrm{Be} \, Z'$, possibly explaining the Atomki anomaly. This because the contributions of A couplings in the transition are proportional to $k/M_{Z'} \ll 1$ (where $k$ is the small momentum of the $Z'$) whereas the V component has a momentum proportionality of $k^3 / M_{Z'}^3$, as explained in \cite{Feng:2016jff}.

In the (purely) V case, the larger values of $(g,\tilde{g})$ conflict with the non-observation of deviations from the SM by neutrino scattering off electrons
(see below, in fact, we detail these experimental requirements on our particular model construction later on). One possibility, explored in \cite{DelleRose:2017xil}, is to employ a 2-Higgs Doublet Model (2HDM), which successfully augments the Yukawa sector such that this condition of gauge invariance is modified by the second Higgs doublet and eventually allows for non-suppressed A couplings. This ensures that the Atomki anomaly can be explained with smaller $g', \tilde g$ gauge couplings which thus alleviate the present experimental constraints.

In this work, we proceed in a different direction. Namely, to allow for A couplings, we consider the possibility of having a family-dependent (non-universal) $U(1)'$. In this case, the Yukawa interaction terms, in  Eq.~(\ref{Yukawa}),  are modified as follows:
\bea
- \mathcal{L}_{Yuk} &=& \Gamma^{u} \dfrac{\chi^{n_{ij}}}{M^{n_{ij}}} \overline{Q}_{L,i}\tilde{H}u_{R,j} 
+ \Gamma^{d} \dfrac{\chi^{l_{ij}}}{M^{l_{ij}}} \overline{Q}_{L,i} H d_{R,j} \nonumber \\
&+&\Gamma^{e} \dfrac{\chi^{m_{ij}}}{M^{m_{ij}}} \overline{L}_{i} H e_{R,j}+ h.c.,
\eea
where the dimension of the non-renormalisable scale $M$ is specified by the $U(1)'$ charges of the involved fields. This procedure can be used to generate fermion masses at tree level or at higher orders\footnote{It may be interesting to investigate whether the same $U(1)'$ symmetry that explains the Atomki anomaly could act as a flavour symmetry and arrange for the observed  fermion mass hierarchy and mixing, however, this is beyond the scope of this paper.} \cite{Froggatt:1978nt}. Therefore, here, we assume $U(1)'$ charges such that the third fermion family Yukawa structure be SM-like, due to more natural $\mathcal{O}(1)$ couplings, while the mass of  first two quark and lepton families can be obtained through some higher order corrections. In fact, various models attempt to explain the smallness of the first two quark and lepton families by a radiative mass generation mechanism, such as \cite{Demir:2005ti}, or by horizontal symmetries \cite{Froggatt:1978nt}. 
Explicitly, we require that the condition in Eq.~(\ref{eq:gauge_invariance})  only holds for the third generation. In short, we choose to impose that the first two generations be flavour universal, but not the third, $z_{i_1}=z_{i_2}$ for $i=\{Q,u_R,d_R,L,e_R\}$. \\
In addition to the aforementioned conditions of gauge invariance of the third generation Yukawa couplings and flavour universality in the first two generations, we now discuss some additional constraints on our charge assignment. \\
Despite working with a low scale, phenomenological approach, we choose to adhere to the chiral anomaly cancellation conditions satisfied by the current fermionic content of the SM in addition to R handed neutrinos. 
The six anomaly conditions are summarised as
	\begin{align}
	\label{eq:anomaly}
	& \sum_i^{3} (2 z_{Q_i} - z_{u_i} - z_{d_i}) = 0 \,,  \\
	&  \sum_i^{3} \, ( 3 z_{Q_i} +  z_{L_i})  = 0 \,,  \\
	& \sum_i^{3} \left( \frac{z_{Q_i}}{6} - \frac{4}{3} z_{u_i} - \frac{z_{d_i}}{3}  +   \frac{z_{{L_i}}}{2} - z_{e_i}\right) = 0 \,,  \\
	& \sum_i^{3} \left( z_{Q_i}^2 - 2 z_{u_i}^2 + z_{d_i}^2      - z_{{L_i}}^2 + z_{e_i}^2 \right) = 0 \,,  \\
	& \sum_i^{3} \left( 6 z_{Q_i}^3 - 3 z_{u_i}^3 - 3 z_{d_i}^3 +  2 z_{{L_i}}^3 - z_{e_i}^3 \right)  + \sum_i^{3} z_{\nu _i}    = 0 \,,  \\
	& \sum_i^{3} \left( 6 z_{Q_i} - 3 z_{u_i} - 3 z_{d_i}  + 2 z_{{L_i}} - z_{e_i} \right) + \sum_i^{3} z_{\nu _i}    = 0 .
	\end{align}
	In order to reduce the number of independent charges, we further impose bounds based on the existing experiments. 
	Firstly, neutrino couplings are strongly constrained by meson decays, such as $K^\pm \rightarrow \pi ^\pm \nu \nu$ \cite{Davoudiasl:2014kua}, and by the electron-neutrino scattering by the TEXONO experiment \cite{Feng:2016ysn,Deniz:2009mu,Bilmis:2015lja,Khan:2016uon}. We thus impose there be no couplings at all for neutrinos to the $Z'$, $C_{\nu,A}=C_{\nu,L}=0$. One finds then the additional requirement that
	\begin{equation}
	z_{L_1}=z_{L_2}=z_{L_3}=-z_{H}.
	\end{equation}
	As stated before, we also require A couplings for the first two generations of quarks to successfully reproduce the Atomki anomaly,
	\begin{align}
	&-z_{Q_{1,2}} -z_H +z_{u_{1,2}} \neq 0 \\
	&-z_{Q_{1,2}} + z_H +z_{d_{1,2}} \neq 0.
	\end{align}
	However, A couplings in the charged lepton sector have stringent constraints from atomic parity violation in caesium (Cs) \cite{Porsev:2009pr}. These can be extracted from the measurement of the effective weak charge $\Delta Q_W$ of the Cs atom:
	\begin{align}
	&\Delta Q_W = \frac{-2\sqrt{2}}{G_F} C_{e,A} 
	\left[C_{u,V} (2 Z + N) + C_{d,V} (Z + 2 N) \right] \nonumber \\
	& \times \left( \frac{0.8}{(17~ {\rm MeV})^2}\right) \lsim 0.71.
	\end{align}
	As, the vector couplings of the $Z'$ to up and down quarks are, in general, non zero, we thus also require that there be no A interaction to the electrons,
	\begin{equation}
	C_{e,A} =0.
	\end{equation}
	This will also help to alleviate bounds from $(g-2)_e$ which are discussed later. For the same reason, we also set the muon A coupling to zero, to avoid increasing the discrepancy between the experimental measurement and the SM prediction of the $(g-2)_\mu$, (discussed further in the paper),
	\begin{equation}
	C_{\mu,A} =0.
	\end{equation}

	With these final constraints, we find that our initial 16 free charges (three generations of $\{z_Q,z_u,z_d,z_L,z_e\}$ and $z_H$) may be expressed as a function of one single parameter. Adjusting this parameter is equivalent to a rescaling of the coupling, so our charge assignment with these constraints is fixed, see Tab. \ref{tab:charges}, and we normalise it with $z_H =1$.
		
	\begin{table}[!t]
		\centering
		\begin{tabular}{|c|c|c|c|c|}
			\hline
			& \multirow{2}{*}{$SU(3)$} & \multirow{2}{*}{$SU(2)$} & \multirow{2}{*}{$U(1)_Y$} & \multirow{2}{*}{$U(1)'$} \\
			&&&&\\ \hline \vspace{-1em}
			&&&&\\
			$Q_{1}$	&  3		&	2	& 1/6	&	$1/3$  \\
			$Q_{2}$	&  3		&	2	& 1/6	&	$1/3$  \\
			$Q_{{3}}$	&  3		&	2	& 1/6	& $1/3$ \\
			$u_{R_{1}}$	&  3		&	1	& 2/3	&	$-2/3$ \\
			$u_{R_{2}}$	&  3		&	1	& 2/3	&	$-2/3$ \\
			$u_{R_{3}}$	&  3		&	1	& 2/3	&	$4/3$  \\
			
			$d_{R_{1}}$	&  3		&	1	& -1/3	&	$4/3$ \\
			$d_{R_{2}}$	&  3		&	1	& -1/3	&	$4/3$\\
			$d_{R_{3}}$	&  3		&	1	& -1/3	&	$-2/3$ \\
			$L_{1}$		&  1		&	2	& -1/2	&	$-1$ \\
			$L_{2}$		&  1		&	2	& -1/2	&	$-1$  \\
			$L_{3}$		&  1		&	2	& -1/2	&	$-1$ \\
			$e_{R_{1}}$	&  1		&	1	& -1		&	$0$ \\
			$e_{R_{2}}$	&  1		&	1	& -1		&	$0$ \\
			$e_{R_{3}}$	&  1		&	1	& -1		&	$-2$ \\
			$H$	&  1		&	2	& 	1/2	&	$1$  \\
			\hline
		\end{tabular}
		\caption{Charge assignment of the SM particles under the family-dependent (non-universal) $U(1)'$. This numerical charge assignment  satisfies the discussed anomaly cancellation conditions, enforces a gauge invariant Yukawa sector of the third generation and family universality in the first two fermion generations as well as  no coupling of the $Z'$ to the all neutrino generations.}
		\label{tab:charges}
	\end{table}
	
	We now discuss the Atomki anomaly requirements and the experimental constraints quantitatively.

The Atomki collaboration \cite{Krasznahorkay:2015iga} has published that the best fit for the mass of the (would be) $Z'$ should be $M_{Z'} = 16.7 \pm 0.35 \textrm{(stat)} \pm 0.5 \textrm{(sys)}$ MeV, corresponding to a ratio of Branching Ratios (BRs),
\begin{align}
\label{eq:Br}
\textrm{Br}  &\equiv {\frac{{\rm BR}(^8{{\rm Be}^*} \to Z' + {^8{\rm Be}})}{{\rm BR}(^8{{\rm Be}^*} \to \gamma + {^8{\rm Be}})} \times {\rm BR}(Z'\to e^+ e^-)}\\ \nonumber
&=5.8 \times 10 ^{-6} ,
\end{align}
with a statistical significance of $\sim 6 \sigma$ \cite{Krasznahorkay:2015iga}. However, the Atomki collaboration has since then pursued further masses and BRs, as mentioned in \cite{Feng:2016ysn}, as a private communication \cite{Krasznahorkay:Private}, though a full analysis of these results has not been presented. Nevertheless, we also write these additional mass and Br values in Tab. \ref{tab:Br}, collecting all the possible solutions to the Atomki anomaly.

The decay width of the excited state of $^8$Be into photons, $\Gamma(^8{{\rm Be}^*} \to \gamma + {^8{\rm Be}})$, is well known $(1.9 \times 10^{-6}$) and in our scenario (due to no L handed neutrino couplings) one has ${\rm BR}(Z' \rightarrow e^+ e^-) =1$. To determine whether a $Z'$ of fixed mass, with specified SM charges under the $U(1)'$ gauge group and kinetic mixing, can satisfy the Atomki anomaly, one must calculate the final piece, $\Gamma(^8{{\rm Be}^*} \to X + {^8{\rm Be}}) \equiv \Gamma$, with upper and lower bounds corresponding to uncertainties in the Nuclear Matrix Elements (NMEs). Specifically, if the ensuing Br lies within upper and lower determinations from NME uncertainties, that particular point is accepted as a successful solution to the Atomki anomaly.

Since we have A couplings we may neglect the (much smaller) V contributions and so we use the results and methodology found in \cite{Kozaczuk:2016nma}. We begin with the partial width $\Gamma$  expressed as
\begin{equation}
\Gamma = \frac{k}{18\pi}\,\left(2+\frac{E_k^2}{M_{Z'}^2}\right) \Big \vert a_n \langle 0 || \sigma^n || 1 \rangle + a_p \langle 0 || \sigma^p || 1 \rangle \Big\vert ^2 ,
\end{equation}
where $E_k ^2 = (E(^8\textrm{Be}^*) - E( {}^{8}\textrm{Be}) )^2 - M_{Z'}^2$, where $E$ refers to the energy of the particular level in the nuclear spectrum. The proton and neutron couplings take values $a_p = (a_0+a_1)/2$ and $a_n=(a_0-a_1)/2$, defined as
\begin{align}
a_0 &= (\Delta u^{(p)}+\Delta d^{(p)})(C_{u,A}+C_{d,A}) + 2 C_{s,A} \Delta s^{(p)} , \label{eq:a0}\\
a_1 &= (\Delta u^{(p)}-\Delta d^{(p)})(C_{u,A}-C_{d,A}) \label{eq:a1},
\end{align}
with coefficients \cite{Bishara:2016hek}
\begin{align} \label{eq:coeff}
\Delta u^{(p)} &= \Delta d^{(n)} ~=~ ~~~0.897(27),\\
\Delta d^{(p)} &= \Delta u^{(n)} ~=~ -0.367(27), \\
\Delta s^{(p)} &= \Delta s^{(n)} ~=~ -0.026(4) . 
\end{align}
Further, the NMEs are \cite{Kozaczuk:2016nma}
\begin{align}
\langle 0^+ \| \sigma_p \| \mathcal{S} \rangle &=  -0.047 (29),\\
\langle 0^+ \| \sigma_n \| \mathcal{S} \rangle &=  -0.132 (33).
\label{eq:nuclear_matrix_elements}
\end{align}
Before evaluating the region of the parameter space explaining the anomalous ${}^8 \textrm{Be} ^*$ transition, though, we ought 
to discuss in more details the various experimental constraints which affect such a low mass and weakly coupled $Z'$.  Firstly, we have not seen such a  $Z'$ in electron beam dump experiments (e.g., SLAC E141) \cite{Riordan:1987aw,Bjorken:2009mm}. Therefore, the $Z'$ has not been produced herein, hence 
	\be 
	C_{e,V}^2 + C_{e,A}^2 < 10^{-17},
	\ee
	or else the $Z'$ has been caught in the dump, hence
	\begin{equation}
	\frac{C_{e,V}^2 + C_{e,A}^2}{{\rm BR}(Z' \to e^+ e^-)} \gsim  3.7 \times 10^{-9}.
	\label{eq:E141_dump}
	\end{equation}
 As the former is not compatible with the Atomki observation, we will consider the latter condition. 
We have also not seen the $Z'$ in the NA64 beam dump experiment {\cite{Banerjee:2018vgk}, which places the (stronger than E141) bound,
 	\begin{equation}
	\frac{C_{e,V}^2 + C_{e,A}^2}{{\rm BR}(Z' \to e^+ e^-)} \gsim  1.6 \times 10^{-8}.
	\label{eq:NA64_dump}
	\end{equation}
 We have not seen a $Z'$ in parity-violating Moller scattering (e.g., SLAC E158) \cite{Anthony:2005pm}. Therefore, the following constraint on the V and A couplings is obtained:
	\begin{equation} 
	\vert C_{e,V}  C_{e,A} \vert \lsim 10^{-8}
	\end{equation}
which is automatically satisfied by our charge assignment.

Also, there are contributions from a $Z'$ to the anomalous magnetic moments of electron and muon \cite{Bennett:2006fi,Blum:2013xva,Lindner:2016bgg}. The one loop contributions $\delta a_{l}$, mediated by a $Z'$, leads to 
	\bea 
	&\delta a_{e} = 7.6 \times 10^{-6} C_{e,V}^2  -3.8 \times 10^{-5} C_{e,A}^2 \\
	&-26 \times 10^{-13} \leq \delta a_{e} \leq 8 \times 10^{-13},\\
	&|\delta a_{\mu}| = |0.009 C_{\mu,V}^2 -C_{\mu,A}^2| \leq  1.6 \times 10^{-9}.
	\eea 
Another constraint is from electron-positron colliders (e.g., KLOE2) \cite{Babusci:2012cr} through $e^+ e^- \to \gamma Z', Z' \to e^+ e^-$. From this process one finds 
	\begin{equation}
	(C_{e,V}^2 + C_{e,A}^2) {\rm BR}(Z' \to e^+ e^-) \lsim 3.7 \times 10^{-7}.
	\end{equation}
There is also a limit due to neutral pion decay, wherein the V couplings of such a light state with quarks are, in general, strongly constrained from $\pi^0 \to Z' +\gamma$
	searches at the NA48/2 experiment \cite{Raggi:2015noa}. The process is proportional to the anomaly factor $ N_\pi = \frac{1}{2} (2 C_{u,V} + C_{d,V})^2$.
	Therefore, one gets the following bound:
	\begin{equation}
	\vert 2 C_{u,V} + C_{d,V})\vert \lsim \frac{0.3 \times 10^{-3}}{\sqrt{{\rm BR}(Z' \to e^+ e^-)}}.
	\end{equation}
Finally, we discuss constraints arising from Flavour Changing Neutral Currents (FCNCs). Despite an initially diagonal charge matrix, as the coupling strength between the first two and third generations differs, rotations into the mass eigenstate will generate off diagonal interactions, in the form of tree level FCNCs. Firstly, we examine $K \rightarrow \pi e^+ e^-$ via  tree level on-shell $Z'$ exchange. Since we have a mass $M_{Z'} \approx 17$ MeV, one does not have contributions to $K \rightarrow \pi \mu^+ \mu^-$. There are strict limits here from LHCb \cite{Aaij:2015dea}, however, there is no sensitivity to our $Z’$ simply because the invariant mass range of $e^+ e^-$ begins from 20 MeV. This is done because the resolution degrades rapidly at small mass due to the background from photon conversion in the detector material. Future measurements may sample from smaller invariant masses, which could act as a discovery tool, or disprove our particular scenario. Next, we turn to $B^0 - \bar{B}^0$ mixing. As a first approximation, we use the results from \cite{Becirevic:2016zri}, but assuming now a light $Z'$, such that the propagator $P \equiv (m_{B} ^2 - M_{Z'}^2)^{-1} \simeq m_{B}^{-2}$, rather than their approximation $P \simeq M_{Z'}^{-2}$. This leads to the condition
\begin{equation}
|g^{L(R)} _{sb}| \lesssim 10^{-6},
\end{equation}
where (upon assuming minimal flavour violation in the quark sector and introducing Cabibbo-Kobayashi-Maskawa (CKM) matrix elements)
\begin{align}
g^L _{sb} &= \Big( V_{\rm CKM} ^T ~\textrm{Diag}(g_{Q_1} ' , g_{Q_1} ' , g_{Q_3} ')~ V_{\rm CKM} \Big) _{23}\\
g^R _{sb} &= \Big( V_{\rm CKM} ^T ~\textrm{Diag}(g_{u_{R_1}} ' , g_{u_{R_1}} ' , g_{u_{R_3}} ')~ V_{\rm CKM} \Big) _{23} 
\end{align} and $g_{i} ' = g' z_{i}$, for $i=\{Q_1 ,~ Q_3 ,~ u_{R_1} ,~u_{R_3} \}$. For our assignment, since it is family universal in the LH quark sector, $g^L _{sb}=0$, and only the RH sector contributes,  $g^R _{sb} \propto V_{tb} V_{ts}$ contributes, which leads to the condition
\begin{equation}
g', \tilde g \lesssim 10^{-4}.
\end{equation}
A similar estimate for the $K - \bar{K}$ mixing yields a less stringent constraint. Despite a smaller propagator suppression (because $m_K < m_B$), the CKM suppression is now much stronger, $\propto V_{td} V_{ts}$ and so one finds the weaker constranint $g', \tilde g \lesssim 10^{-3}$.
In the scope of this paper, we do not perform a full flavour analysis of the $(B - \bar{B})$ and $(K - \bar{K})$ mixing, but leave this as an approximate requirement.

We now present the results for our particular charge assignment shown in Tab. \ref{tab:charges}, consistent with all of the aforementioned experimental constraints\footnote{Other charge assignments are also possible by relaxing the conditions we impose.}. In Fig. \ref{fig:1HDM_ZPHI_0.5_ZQ3_-1Region}, we plot the allowed parameters in the space of the $U(1)'$ gauge coupling, $g'$, and the gauge-kinetic mixing strength, $\tilde{g}$. Regions which can satisfy the results of the Atomki experiment are shown in red, purple and green, corresponding to the three different mass solutions of 16.7, 17.3 and 17.6 MeV, respectively. One can see that these bands overlap in places. The bands are independent of $\tilde{g}$ because the Atomki anomaly depends on axial couplings, which are independent of $\tilde{g}$ and $Br(Z' \rightarrow e^+ e^-)=1$ for all $(g',\tilde{g})$. Also shown are the requirements from $(g-2)_e$ (allowed regions are inside the two dotted line boundary, shaded in blue), $(g-2)_\mu$ (allowed regions are inside the two dashed line boundary), and the electron beam dump experiment, NA64 (allowed regions are outside the two solid lines, arising from Eq. \ref{eq:NA64_dump} (ie not at $\tilde{g}=0$ for small $g'$) and also shaded in blue). 

The other constraints (electron positron collider (KLOE2), Moller scattering (E158), pion decay (NA48/2) and atomic parity violation of Cs) are  satisfied by all regions of the shown parameter space and so are not displayed on the plot. The constraint from E141 is strictly less constraining than NA64 and so also not displayed on the plot.

The total allowed parameter space is thus in the dark blue shaded regions, on top of the solutions to the Atomki anomaly for all three masses, shaded in red, purple and green.

\begin{table}[!t]
\begin{tabular}{c | c}
	$M_{Z'} ~(\textrm{MeV}) $ & Br \\ \hline \vspace{-1em}
	&\\ 
	16.7 & $5.8 \times 10^{-6}$ \\
	17.3 & $2.3 \times 10^{-6}$ \\
	17.6 & $5.0 \times 10^{-7}$
\end{tabular}
\caption{Solutions to the Atomki anomaly, with best fit mass value (16.7 MeV) from \cite{Krasznahorkay:2015iga} and subsequent alternative masses (17.3 MeV and 17.6 MeV) from \cite{Feng:2016ysn} along with the corresponding ratio of BRs, Br, as defined in Eq. (\ref{eq:Br}).}
\label{tab:Br}
\end{table}

\begin{figure}[h]
\centering
\includegraphics[width=1.0\linewidth]{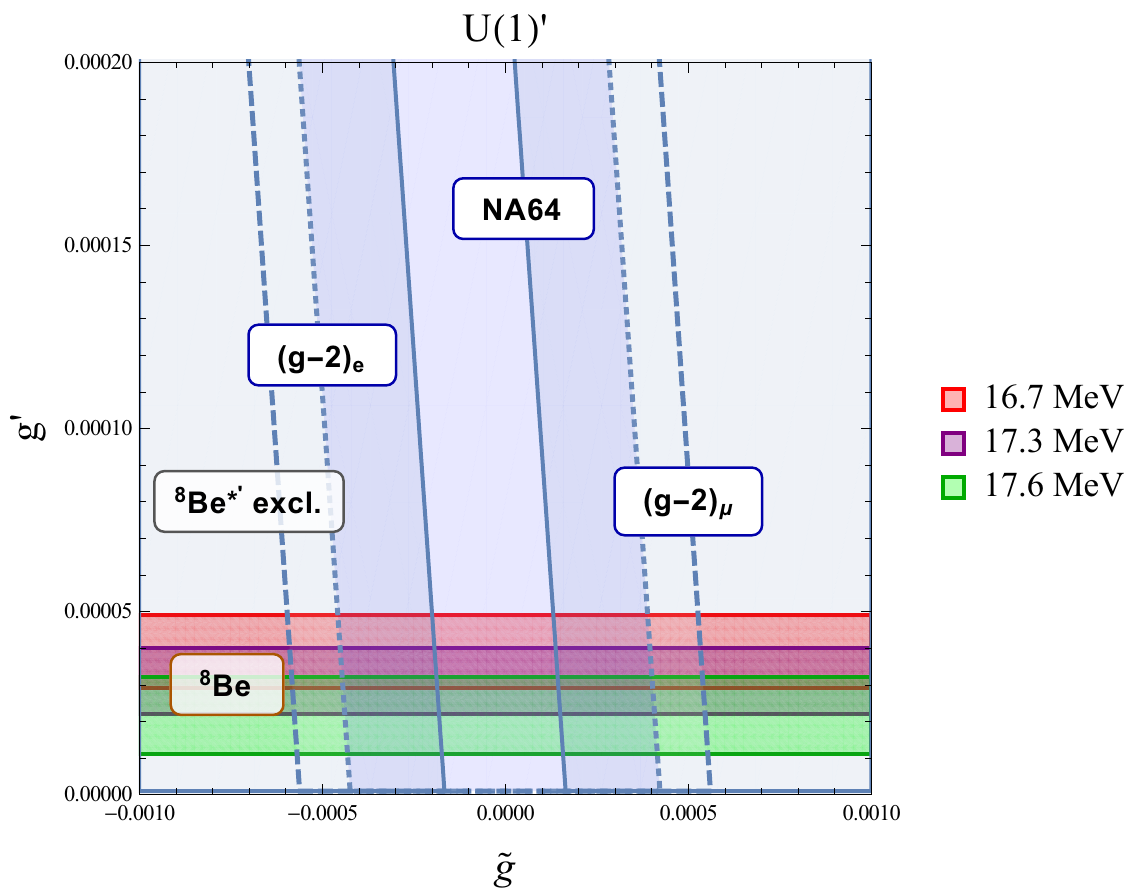}
\caption{Allowed parameter space mapped on the $(g',\tilde{g})$ plane 
explaining the anomalous ${}^8 \textrm{Be}^*$ decay for $Z'$ solutions with mass 16.7 (red), 17.3 (purple) and 17.6 (green) MeV. The white regions are excluded by the non-observation of the same anomaly in the ${}^8 \textrm{Be}^{*'}$ transition. Also shown are the constraints from $(g-2)_\mu$, to be within the two dashed lines; $(g-2)_e$, to be inside the two dotted lines (shaded in blue) and the electron beam dump experiment, NA64, to be in the shaded blue region outside the two solid lines. The surviving parameter space lies at small positive and negative $\tilde{g}$ (though not at $\tilde{g}=0$), inside the dark shaded blue region which overlaps the Atomki anomaly solutions.}
\label{fig:1HDM_ZPHI_0.5_ZQ3_-1Region}
\end{figure}

Fig.~\ref{fig:BRPlotDensity_high} shows the quantity BR, defined in Eq. (\ref{eq:Br}), for given values of $M_{Z'}$. For each mass value a scan has been done over the allowed parameter space in $(g',\tilde{g})$ from Fig.~\ref{fig:1HDM_ZPHI_0.5_ZQ3_-1Region} which may explain the Atomki anomaly. There is no fixed BR for each $\{M_{Z'},~g',~\tilde{g} \}$, but a range due to the uncertainties in the NMEs of Eq. (\ref{eq:nuclear_matrix_elements}). One finds that the lower limit of BR is always smaller than that of the Atomki anomaly. Therefore, only the upper limit of it is of interest and only the corresponding values are plotted following the scan (in blue). The Atomki collaboration measurements are also shown (in orange). Upper limit BR  points which lie above the Atomki results consequently provide valid explanations of the anomaly.
For a given mass, one can see the trend to have a larger density of upper BR bounds at smaller values of it. Furthermore, the largest upper bound decreases with heavier $Z'$ masses.
For the 16.7 MeV mass point, there are many points which lie below the Atomki solution and so are not valid descriptions to explain the anomaly. Yet, there are plenty of valid points above it too. However, for 17.3 MeV and particularly 17.6 MeV, the majority of points lie above the required BR and so are all acceptable solutions.
The combination of these two effects motivate why heavier $M_{Z'}$ values have a larger range of solutions (ie a thicker green (17.6 MeV) than red (16.7 MeV) band in Fig. \ref{fig:1HDM_ZPHI_0.5_ZQ3_-1Region}.

\begin{figure}[h]
\centering
\includegraphics[width=1.0\linewidth]{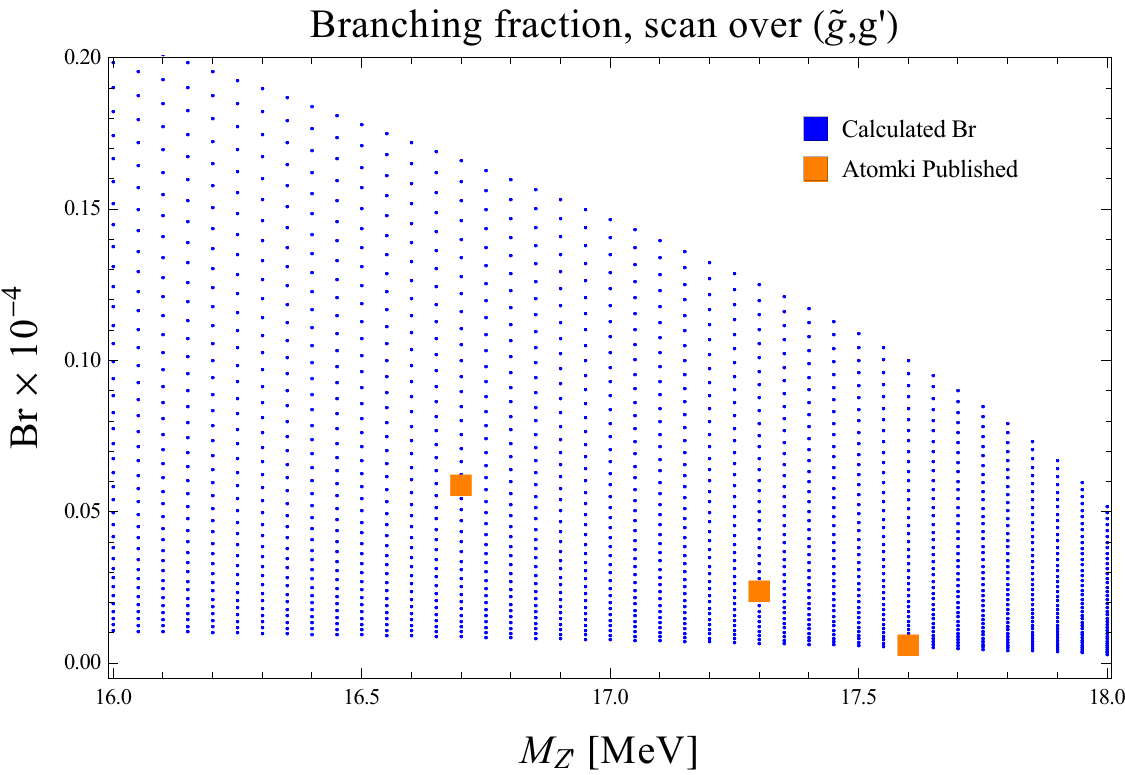}
\caption{Values of the upper limit Br (lower limits are smaller than the scale of the plot), as defined in Eq. (\ref{eq:Br}), versus the mass of the $Z'$ obtained scanning over the allowed parameter space in $(g',\tilde{g}$, obtained from Fig.~\ref{fig:1HDM_ZPHI_0.5_ZQ3_-1Region} for each mass step taken (in blue). The Atomki collaboration solutions are also shown (in orange).}
\label{fig:BRPlotDensity_high}
\end{figure}

{{In conclusion, with the assumption that the first two families of SM quark and lepton masses are generated by some high scale physics, unlike those of the third one which stem from a SM Higgs mechanism supplemented by an additional $U(1)'$ (broken) group, yielding a very light $Z'$ state,  
we have found a family-dependent (non-universal) charge assignment which can successfully accommodate the Atomki anomaly, in addition to all other experimental constraints on such a low scale physics. This happens for a range of $Z'$ masses (and corresponding decay rates), including the best fit of $M_{Z'}=16.7$ MeV as well as other two published values, $17.3$ and $17.6$ MeV, over the coupling ranges $g' \sim 10^{-5}$ and $1 \times 10^{-5} \lesssim | \tilde{g} | \lesssim 5 \times 10^{-5}$ for the gauge and kinetic mixing couplings, respectively, regulating the $Z'$ interactions with SM fermions.}

\section*{Acknowledgements}
The work of LDR and SM is supported in part by the NExT Institute. SM also acknowledges partial financial contributions from the STFC Consolidated Grant ST/L000296/1. Furthermore, the work of LDR has been supported by the STFC/COFUND Rutherford International Fellowship Programme (RIFP). SJDK and SK have received support under the H2020-MSCA grant agreements InvisiblesPlus (RISE) No. 690575 and Elusives (ITN) No. 674896. In addition SK was partially supported by the STDF project 13858. All authors acknowledge support under the H2020-MSCA grant  agreement NonMinimalHiggs (RISE)
 No. 645722.


\begin{thebibliography}{99}

\bibitem{Gulyas:2015mia} 
  J.~Guly\'as {\it et al.},
  Nucl.\ Instrum.\ Meth.\ A {\bf 808} (2016) 21
  [arXiv:1504.00489 [nucl-ex]].

	\bibitem{Krasznahorkay:2015iga}
	A.~J.~Krasznahorkay {\it et al.},
	Phys.\ Rev.\ Lett.\  {\bf 116} (2016) no.4,  042501
	[arXiv:1504.01527 [nucl-ex]].

\bibitem{Krasznahorkay:2017gwn} 
  A.~J.~Krasznahorkay {\it et al.},
  EPJ Web Conf.\  {\bf 142} (2017) 01019.

\bibitem{Krasznahorkay:2017bwh} 
  A.~J.~Krasznahorkay {\it et al.},
  PoS BORMIO {\bf 2017}  (2017) 036.

\bibitem{Krasznahorkay:2017qfd} 
  A.~J.~Krasznahorkay {\it et al.},
  EPJ Web Conf.\  {\bf 137} (2017) 08010.

\bibitem{Krasznahorkay:2018snd} 
  A.~J.~Krasznahorkay {\it et al.},
  J.\ Phys.\ Conf.\ Ser.\  {\bf 1056} (2018) no. 1, 012028. 

	\bibitem{Feng:2016ysn}
	J.~L.~Feng, B.~Fornal, I.~Galon, S.~Gardner, J.~Smolinsky, T.~M.~P.~Tait and P.~Tanedo,
	Phys.\ Rev.\ D {\bf 95} (2017) no.3,  035017
	[arXiv:1608.03591 [hep-ph]].
	
	\bibitem{Feng:2016jff}
	J.~L.~Feng, B.~Fornal, I.~Galon, S.~Gardner, J.~Smolinsky, T.~M.~P.~Tait and P.~Tanedo,
	Phys.\ Rev.\ Lett.\  {\bf 117} (2016) no.7,  071803
	[arXiv:1604.07411 [hep-ph]].
	
	\bibitem{Gu:2016ege}
	P.~H.~Gu and X.~G.~He,
	Nucl.\ Phys.\ B {\bf 919} (2017) 209
	[arXiv:1606.05171 [hep-ph]].
	
	
	\bibitem{Chen:2016dhm}
	L.~B.~Chen, Y.~Liang and C.~F.~Qiao,
	arXiv:1607.03970 [hep-ph].
	
	\bibitem{Liang:2016ffe}
	Y.~Liang, L.~B.~Chen and C.~F.~Qiao,
	Chin.\ Phys.\ C {\bf 41} (2017) no.6,  063105
	[arXiv:1607.08309 [hep-ph]].
	
	\bibitem{Jia:2016uxs}
	L.~B.~Jia and X.~Q.~Li,
	Eur.\ Phys.\ J.\ C {\bf 76} (2016) no.12,  706
	[arXiv:1608.05443 [hep-ph]].
	
	\bibitem{Kitahara:2016zyb}
	T.~Kitahara and Y.~Yamamoto,
	Phys.\ Rev.\ D {\bf 95} (2017) no.1,  015008
	[arXiv:1609.01605 [hep-ph]].
	
	\bibitem{Chen:2016tdz}
	C.~S.~Chen, G.~L.~Lin, Y.~H.~Lin and F.~Xu,
	Int.\ J.\ Mod.\ Phys.\ A {\bf 32} (2017) no.31,  1750178
	[arXiv:1609.07198 [hep-ph]].
	
	\bibitem{Seto:2016pks}
	O.~Seto and T.~Shimomura,
	Phys.\ Rev.\ D {\bf 95} (2017) no.9,  095032
	[arXiv:1610.08112 [hep-ph]].
	
	\bibitem{Neves:2016ugb}
	M.~J.~Neves and J.~A.~Helayël-Neto,
	arXiv:1611.07974 [hep-ph].
	
	\bibitem{Chiang:2016cyf}
	C.~W.~Chiang and P.~Y.~Tseng,
	Phys.\ Lett.\ B {\bf 767} (2017) 289
	[arXiv:1612.06985 [hep-ph]].
	
\bibitem{Ellwanger:2016wfe}
  U.~Ellwanger and S.~Moretti,
  JHEP {\bf 1611} (2016) 039
  [arXiv:1609.01669 [hep-ph]].

	\bibitem{DelleRose:2017xil}
	L.~Delle Rose, S.~Khalil and S.~Moretti,
	Phys.\ Rev.\ D {\bf 96} (2017) no.11,  115024
	[arXiv:1704.03436 [hep-ph]].
	
	\bibitem{Coriano:2015sea}
	C.~Coriano, L.~Delle Rose and C.~Marzo,
	JHEP {\bf 1602} (2016) 135
	[arXiv:1510.02379 [hep-ph]].
	
\bibitem{Froggatt:1978nt}
  C.~D.~Froggatt and H.~B.~Nielsen,
  Nucl.\ Phys.\ B {\bf 147} (1979) 277.
	
	
	\bibitem{Demir:2005ti}
	D.~A.~Demir, G.~L.~Kane and T.~T.~Wang,
	Phys.\ Rev.\ D {\bf 72} (2005) 015012
	[hep-ph/0503290].
	
	\bibitem{Davoudiasl:2014kua}
	H.~Davoudiasl, H.~S.~Lee and W.~J.~Marciano,
	Phys.\ Rev.\ D {\bf 89} (2014) no.9,  095006
	[arXiv:1402.3620 [hep-ph]].
	
	\bibitem{Deniz:2009mu}
	M.~Deniz {\it et al.} [TEXONO Collaboration],
	Phys.\ Rev.\ D {\bf 81} (2010) 072001
	[arXiv:0911.1597 [hep-ex]].
	
	\bibitem{Bilmis:2015lja}
	S.~Bilmis, I.~Turan, T.~M.~Aliev, M.~Deniz, L.~Singh and H.~T.~Wong,
	Phys.\ Rev.\ D {\bf 92} (2015) no.3,  033009
	[arXiv:1502.07763 [hep-ph]].
	
	\bibitem{Khan:2016uon}
	A.~N.~Khan,
	Phys.\ Rev.\ D {\bf 93} (2016) no.9,  093019
	[arXiv:1605.09284 [hep-ph]].
	
	\bibitem{Krasznahorkay:Private}
	A. J. Krasznahorkay, private communication, 2016.
	
	
	
	
%
%
%
%
	
	
	\bibitem{Kozaczuk:2016nma}
	J.~Kozaczuk, D.~E.~Morrissey and S.~R.~Stroberg,
	Phys.\ Rev.\ D {\bf 95} (2017) no.11,  115024
	[arXiv:1612.01525 [hep-ph]].
	
	\bibitem{Bishara:2016hek}
	F.~Bishara, J.~Brod, B.~Grinstein and J.~Zupan,
	JCAP {\bf 1702} (2017) no.02,  009
	[arXiv:1611.00368 [hep-ph]].
	
	\bibitem{Riordan:1987aw}
	E.~M.~Riordan {\it et al.},
	Phys.\ Rev.\ Lett.\  {\bf 59} (1987) 755.
	
	\bibitem{Bjorken:2009mm}
	J.~D.~Bjorken, R.~Essig, P.~Schuster and N.~Toro,
	Phys.\ Rev.\ D {\bf 80} (2009) 075018
	[arXiv:0906.0580 [hep-ph]].
	
	\bibitem{Banerjee:2018vgk}
	 D.~Banerjee {\it et al.} [NA64 Collaboration],
	 Phys.\ Rev.\ Lett.\  {\bf 120} (2018) no.23,  231802
	 doi:10.1103/PhysRevLett.120.231802
	 [arXiv:1803.07748 [hep-ex]].
	
	\bibitem{Anthony:2005pm}
	P.~L.~Anthony {\it et al.} [SLAC E158 Collaboration],
	Phys.\ Rev.\ Lett.\  {\bf 95} (2005) 081601
	[hep-ex/0504049].
	
	\bibitem{Bennett:2006fi}
	G.~W.~Bennett {\it et al.} [Muon g-2 Collaboration],
	Phys.\ Rev.\ D {\bf 73} (2006) 072003
	[hep-ex/0602035].
	
	\bibitem{Blum:2013xva}
	T.~Blum, A.~Denig, I.~Logashenko, E.~de Rafael, B.~Lee Roberts, T.~Teubner and G.~Venanzoni,
	arXiv:1311.2198 [hep-ph].
	
	\bibitem{Lindner:2016bgg}
	M.~Lindner, M.~Platscher and F.~S.~Queiroz,
	Phys.\ Rept.\  {\bf 731} (2018) 1
	[arXiv:1610.06587 [hep-ph]].
	
	
	
	
	\bibitem{Babusci:2012cr}
	D.~Babusci {\it et al.} [KLOE-2 Collaboration],
	Phys.\ Lett.\ B {\bf 720} (2013) 111
	[arXiv:1210.3927 [hep-ex]].
	
	\bibitem{Raggi:2015noa}
	M.~Raggi [NA48/2 Collaboration],
	Nuovo Cim.\ C {\bf 38} (2016) no.4,  132
	[arXiv:1508.01307 [hep-ex]].
	
	\bibitem{Aaij:2015dea}
	R.~Aaij {\it et al.} [LHCb Collaboration],
	JHEP {\bf 1504} (2015) 064
	doi:10.1007/JHEP04(2015)064
	[arXiv:1501.03038 [hep-ex]].
	
	\bibitem{Becirevic:2016zri}
	D.~Bečirević, O.~Sumensari and R.~Zukanovich Funchal,
	Eur.\ Phys.\ J.\ C {\bf 76} (2016) no.3,  134
	doi:10.1140/epjc/s10052-016-3985-0
	[arXiv:1602.00881 [hep-ph]].
	
		\bibitem{Porsev:2009pr}
		S.~G.~Porsev, K.~Beloy and A.~Derevianko,
		Phys.\ Rev.\ Lett.\  {\bf 102} (2009) 181601
		[arXiv:0902.0335 [hep-ph]].
\end{thebibliography}
\end{document}